# A Neutron Spectrometry for Fusion Plasmas

Iskender Atilla REYHANCAN

Istanbul Technical University, Energy Institute, Ayazaga Campus, Maslak, Istanbul, Turkiye

**Abstract**

It is very important to measure the plasma performances of fusion reactors such as fuel ion density, plasma temperature, fusion power, fast neutron energy, and neutron flux. Various diagnostic systems are needed to measure fast neutron properties in fusion plasmas. Generally, various types of neutron cameras and monitors are used. These systems are sensitive to neutron as well as gamma-ray radiation. Therefore, monitor systems that distinguish between neutron and gamma rays should be used. In this study, the NE-213 liquid scintillation detector was selected, and Pulse Shape Discriminator (PSD) NIM unit was used for the neutron/gamma-ray discrimination. True spectrum was found by unfolding the spectrum obtained by the reaction of neutrons to the protons in the detector with the FERD-PC computer code. In this study, the energy spectrum of fast neutrons produced in a neutron generator was measured by the unfolding method. Fast neutrons are obtained via the deuteron-tritium ($^2$H+$^3$H $\rightarrow$ $^4$He+n) reaction and have a single energy of about 14 MeV.

1. Introduction

Fast neutrons formed in fusion reactors generally consist of Deuteron+Deuteron (DD) and Deuteron+Tritium (DT) reactions. In the DT tokamak reactor, detectors are used to measure the total fusion energy and energy spectra of these neutrons in the reactor [1]. Thus, it is possible to get accurate information about the operating status of the reactor. A kind of neutron detection systems are used in the fusion reactors like ITER, and JET. These systems consist of neutron scintillator detectors. As a result of various reactions of neutrons around and inside the detector, gamma rays are formed. These gamma rays and neutrons create their electrical signals in neutron detectors. There are many techniques for making neutron/gamma-ray separation in these detectors. Some of them are the time-of-flight (TOF) technique and the spectrum unfolding technique. While long distances and highly complex systems are required for the TOF technique, they are not needed much in the unfolding technique [2].

In this work, the events caused by neutrons in the NE-213 liquid scintillation detector, the general structure of the detector, the extraction of gamma rays caused by neutrons with the Pulse Shape Discriminator (PSD) [3] the calibration of the detector and other tests were performed. In addition, the solution to the unfolding problem of the folded recoil proton spectrum in the detector and a general introduction to the computer code (FERD-PC) [4] were given in the manuscript.

In this paper, energy calibration of the detector was performed with various standard gamma-ray sources. Next, tests were performed with the known $^{241}$Am-Be isotopic neutron source. Afterward, the study was carried out with 14 MeV neutrons produced by the d+T→n+$^{4}$He reaction using the SAMES J-15 neutron generator. Here, the deuterons were accelerated with 150 kV acceleration potential and the Tritium target was bombarded with them. The 14 MeV neutrons formed due to the reaction were used in this study. Thus, monitoring the production of 14 MeV neutrons formed in a fusion reactor will be possible.

2. Description of the system

*2.1. Neutron Detector (NE213)*

A liquid scintillation detector was used to measure the neutron energy spectrum. This detector is NE-213 type and is connected to a phototube and a base. The phototube has two different types of output. These are the anode and the dynode outputs. While the anode output gives fast-timing signals, the dynode output gives slow-timing signals. Neutrons cause a gamma-ray emission as a result of various interactions with the environmental material. At the same time, since the NE-213 liquid scintillation detector is sensitive to gamma rays, there are gamma-ray signals as well as neutrons in the fast-timing signals received from the anode of the phototube. Neutrons typically produce longer and slower scintillation pulses, while gamma rays produce shorter and faster scintillation pulses. This difference in pulse shape can be exploited to discriminate between neutron and gamma-ray events. To distinguish these signals from each other, a "Pulse Shape Discriminator" (PSD) [3] should be used.  The NE-213 liquid scintillation detector that was used as a neutron detector was connected to the PSD NIM unit. The detector is 5 cm in diameter and 5 cm in height. An RCA 8575 phototube and a base (Ortec Inc. Model 216) [5] were connected to this detector. A high voltage of -1700 V was applied to the base during this study.

It's important to note that discrimination techniques may have limitations and there may be cases where discrimination between neutron and gamma-ray events is challenging or not possible with high accuracy. The effectiveness of discrimination techniques in a NE213 liquid scintillator detector may depend on various factors such as the energy range of the particles, the design of the detector, the specific application, and the experimental conditions.

### 2.2. Pulse Shape Discriminator (PSD)

PSD Unit [3] is based on development at Technical Universität München. The technique used is similar to the "zero-crossing" circuit [6]. The most important advantage of this technique is that the system can be controlled quickly with minimum adjustment and operates in an extended dynamic range. The unit provides optimum pulse shape separation for liquid scintillation counters. However, the applications are not limited to n/γ-ray separation, but the model 2160A can also be used for particle separation with inorganic scintillators, Phoswitches thick surface barrier detectors, and proportional counters. This unit is used with a "Constant Fraction Discriminator" (CFD) (Ortec Inc., Model 584) [7]. The constant fraction discriminator used with this unit is the most important unit used to process timing signals. In the constant fraction technique, a part of the signal coming into the circuit is delayed. The amplitude of the delayed signal is varied and subtracted from the input signal. Thus, a bipolar signal is generated, the zero-crossing point of this signal is detected, and an output pulse is produced. The delay time is determined by connecting the appropriate length of cable between the two delay ports on the front of the unit. This process is optimized for each application. This optimization depends on the nominal width and rising time of the input signal. The connection between these two units is shown in Fig. 1 in detail. PSD can distinguish according to the shape of the pulses at the anode output of the phototube. For this, the "rise time" of the pulses is used. That's why another name for this unit is "rise-time discriminator". Depending on the effect of neutron and gamma rays in the scintillation detector, the shape of the pulse at the anode output changes. Thus, two different types of radiation can be separated from each other. For this reason, the background caused by the gamma-ray can be separated from the neutron signals. One of the CFD's time outputs is used as a "strobe" signal to locate the zero-crossing point of gamma and neutron signals in the PSD. Thus, gamma rays and neutrons are separated from each other by the strobe signal. What remains are neutron timing pulses. The linear signals from the $10^{th}$ dynode (dynode output of the phototube) are then gated with separated timing pulses (from the PSD side). Thus, recoiling proton signals belonging to neutrons in the desired energy region is obtained. These signals are converted into a pulse height spectrum with Analog

Digital Converter (ADC#1) (Canberra Industries, Inc., Model 8075) [8] and Multi-Channel Analyzer (MCA) (Canberra Industries, Inc., 1986b) [9].

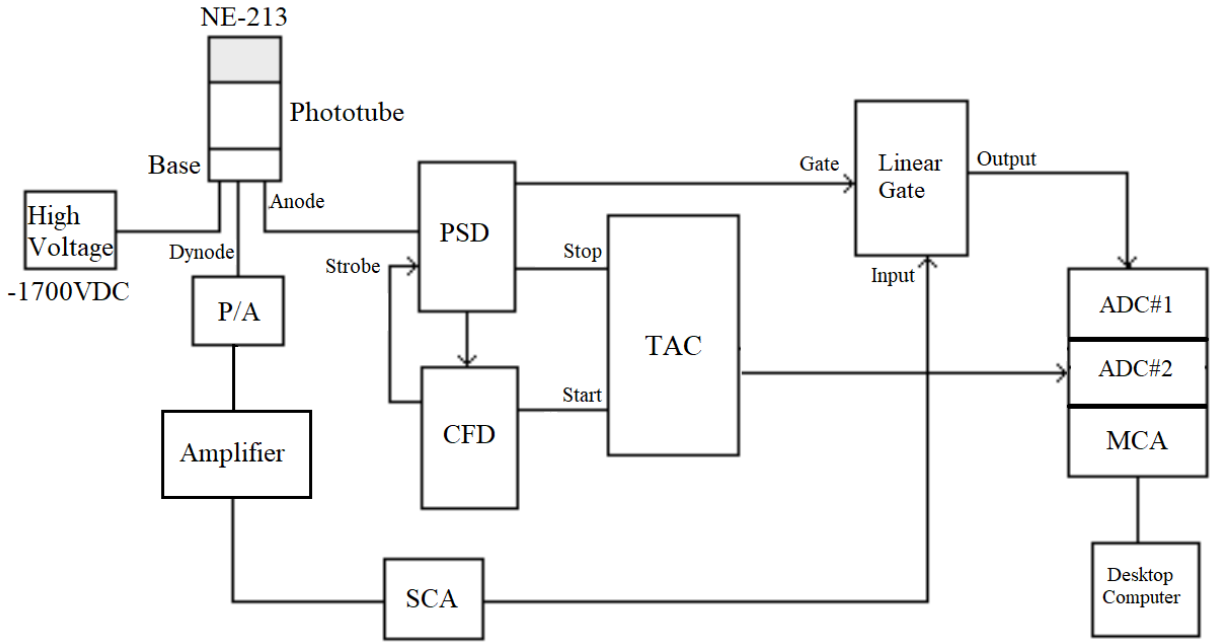

Figure 1. Block diagram of the neutron detection system with the NE-213 detector

### 3. Performance of the PSD Unit

A series of experiments were performed to show how the Pulse Shape Discrimination changes neutron/gamma-ray separation over various energy ranges. Spectra showing neutron and gamma-ray separation were obtained by connecting the TAC output to the ADC#2 unit. For this purpose, the neutron source of $^{241}$Am-Be (30mCi) was placed near the NE213 scintillator, and the neutron/gamma-ray separation became more pronounced in increasing various gating energies (Figure 2).

As can be seen in Figure 3, the ratios between the gamma-ray peak and valley were found as 1/197, and 1/220 in neutron/gamma-ray spectra taken at various threshold energies (0.24 MeV and 0.5 MeV respectively) of the CFD. As it can be understood from here, the distinction becomes more pronounced as the threshold energy is increased in CFD. For this reason, the threshold value was taken as 0.5 MeV during the experiment. At the same time, the "Figure of Merit" (FOM) values of each spectrum were calculated [2].

$$FOM = \frac{X}{W_n + W_\gamma}$$

Where X is the separation between neutron and gamma-ray peaks, $W_n$ and $W_\gamma$ are FWHM of neutron and gamma-ray peaks respectively.

Table 1. FOM values

| Threshold energy | FOM |
|---|---|
| 0.24 MeV | 1.15 |
| 0.5 MeV | 1.54 |

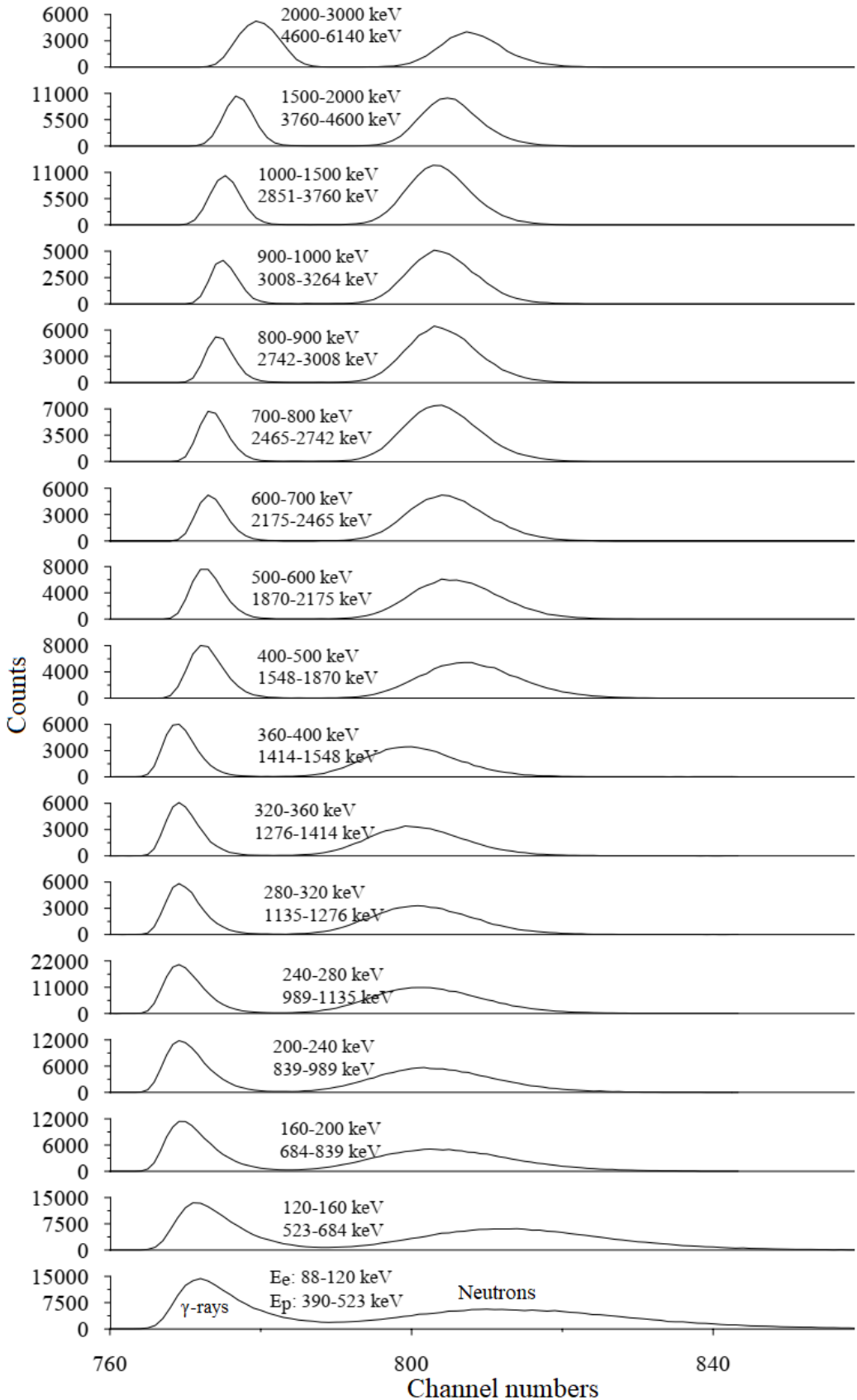

Figure 2. Performance of the Pulse Shape Discriminator (PSD)

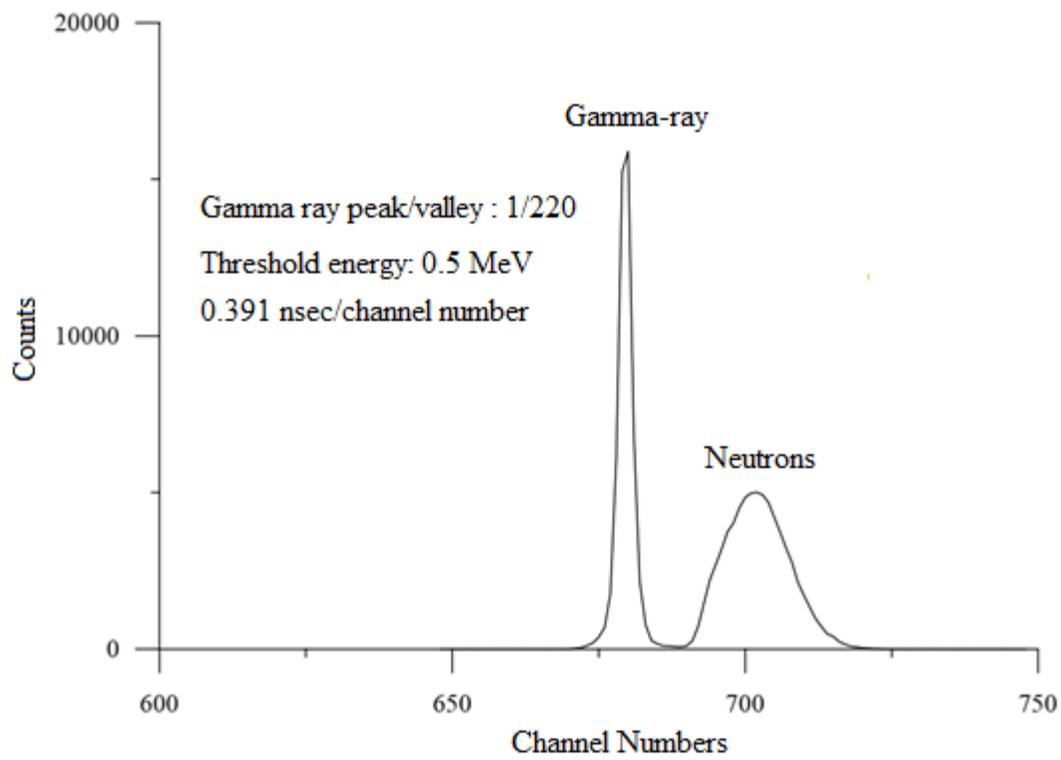

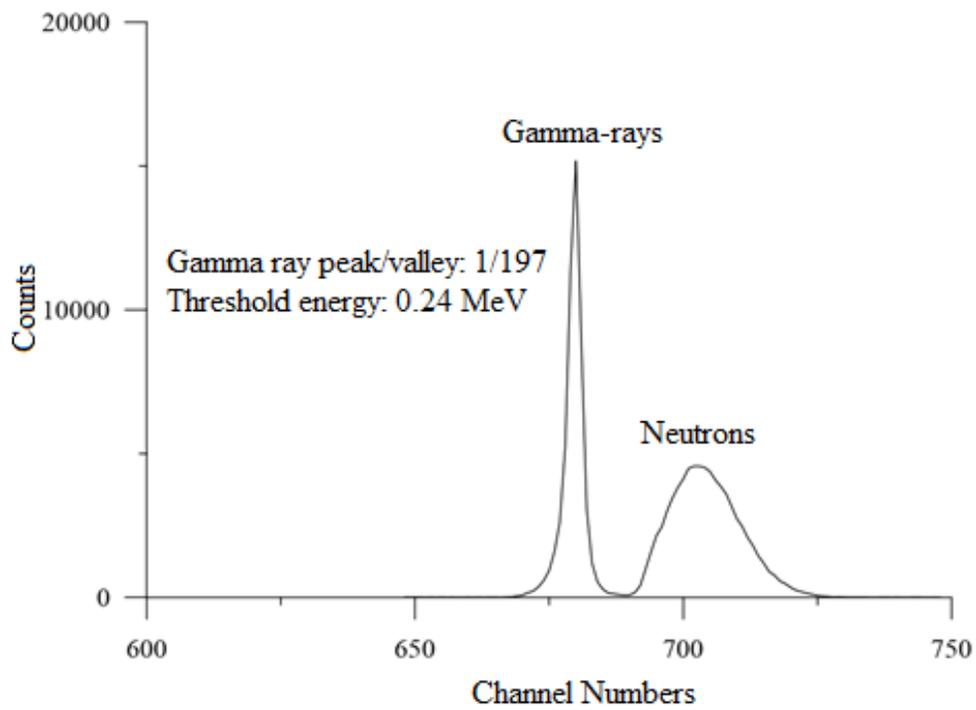

Figure 3. Neutron-gamma ray separation at various threshold energies with PSD

## 4. Calibration of the Neutron Detector

For the energy calibration of the NE-213 liquid scintillation detector, gamma-ray sources of known energies were used. A "standard light unit" was defined for the pulse height distributions of gamma-ray sources measured [10]. A standard light unit based on the measured gamma-ray radiation pulse-height distribution proposed by Verbinski et al. was adopted in the present study. 1.13 of the half-height of the "Compton edge" of the gamma rays with 1.275 MeV energy of the $^{22}$Na gamma-ray source was taken as 1 light unit (1 L) [10]. In addition, the analyzer (MCA) used in the experiment must be zeroed for absolute pulse height calibration. Thus, MCA used for the study must be properly zeroed for absolute pulse height calibration. It became necessary to determine the zero intercept for each spectrum. This was achieved by using standard gamma-ray sources and calibrating each spectrum in terms of electron energy. For photon energies up to 3 MeV Compton scattering dominates in the NE213 scintillator and recoil electrons are produced by monoenergetic gamma radiation having energies up to seven standard gamma-ray calibration sources were used for energy calibration. There is a relationship between the energies of gamma rays and the energies of the ejected electrons (for the scattering angle 180° in Compton Scattering).

$$E_e = \frac{2E_\gamma^2}{(0.511 + 2E_\gamma)}$$

Compton edge will be used for energy calibration. There are many approaches to this. Among them, the approach of Beghian et al. [11] was used. Accordingly, 2/3 of the peak of the sharp edge of the Compton distribution best determines the maximum ejected electron energy. Two-thirds of the Compton edges ($E_c$) of the seven gamma-ray sources were used for energy calibration and determination of the intercept point. Channel Number (CN) values of Compton edges corresponding to each gamma-ray energy are given in Table 2. The spectra of the sources used for calibration are shown in Figure 4. Here, the gamma-ray source (12C*, 4.42 MeV) of the last spectrum is formed by the $^9$Be($\alpha$,n)$^{12}$C* reaction in the $^{241}$Am-Be source. To count the gamma rays at this energy, the gamma rays were separated from the neutrons by using the PSD, and only the gamma spectrum was obtained thanks to the linear gate circuit. After the Compton edges of the spectra taken for each source were determined, a fit was made using the least squares method to a linear line according to their gamma-ray energies (Figure 5). The least-square fitted line yields the relation, 1 light unit = 1.25 MeV, by taking the maximum electron energy at 2/3, the height of the Compton edge.

Table 2. Gamma-ray sources used for energy calibration of the NE213 neutron detector

| Gamma-ray source | Half-life | $E_\gamma$ (keV) | Compton energy (keV) | $E_c$ Ch# | FWHM Ch# |
|---|---|---|---|---|---|
| $^{22}$Na | 950.4 days | 1275 | 1062 | 98 | 21 |
| $^{54}$Mn | 312.5 days | 835 | 639 | 52 | 11 |
| $^{65}$Zn | 244 days | 1115 | 908 | 81 | 19 |
| $^{228}$Th | 1.91 year | 2615 | 2382 | 238 | 36 |
| $^{60}$Co | 5.271 years | 1334 | 1118 | 104 | 22 |
| $^{137}$Cs | 30 years | 662 | 477 | 34 | 9 |
| $^{12}$C* | 431 years | 4420 | 4178 | 429 | 57 |

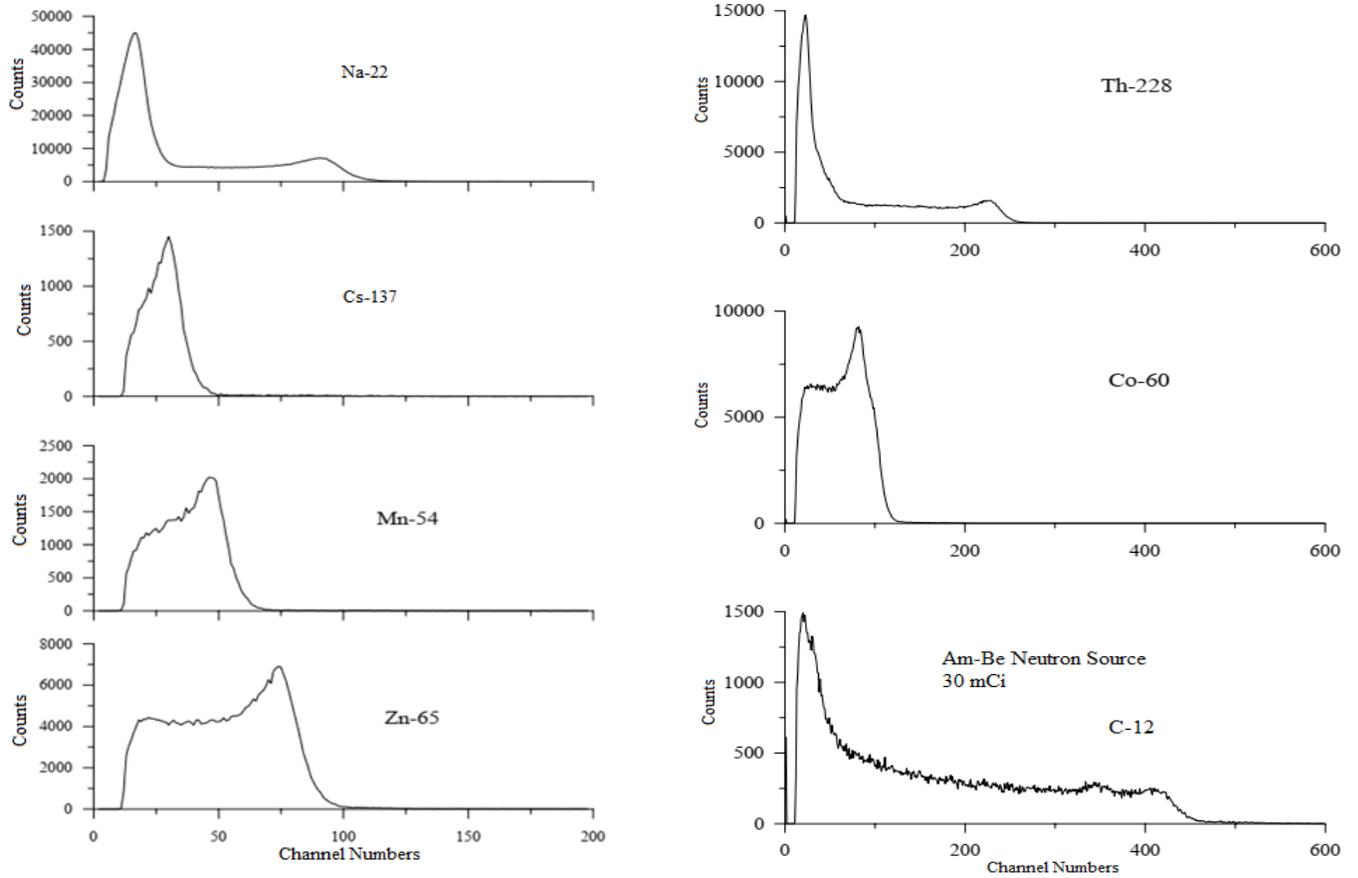

Figure 4. Energy spectra of gamma-ray sources

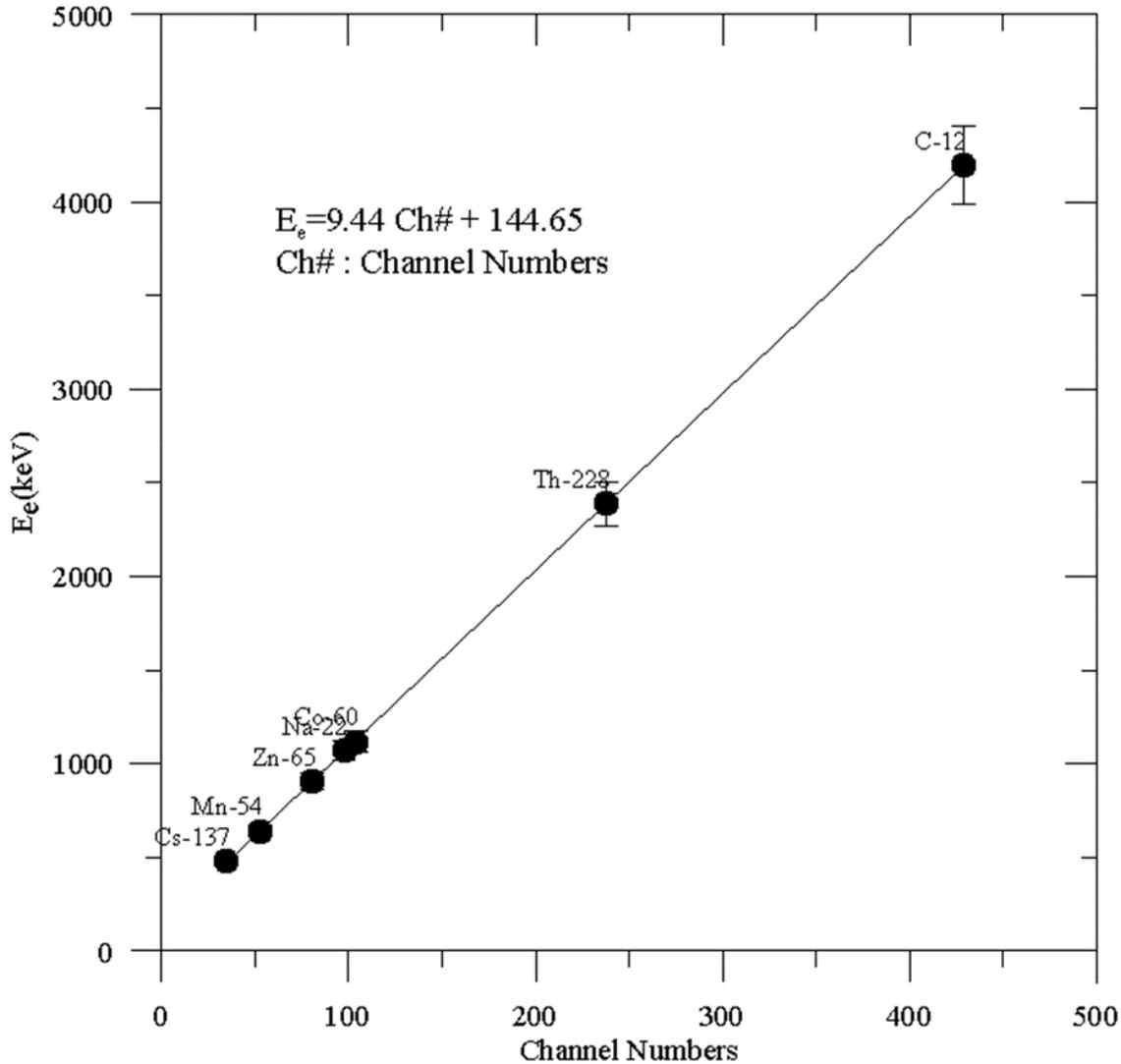

Figure 5. Energy calibration of the detection system

## 5. Resolution of the Neutron Detector

Gamma sources used in energy calibration were used to determine the energy separation power (resolution) of the NE213 detector. For this, the method of Dietze and Klein [12] was followed. The Compton edge of each source was fitted to a Gaussian function and "half-height full width" (FWHM) values were found. The energy equivalents in MeV and from there (1L = 1.25 MeV) L and ΔL values were calculated with the calibration function. If R is represented by resolution, an R-value for each $L$ was found from the expression $R=\Delta L/L$, and a, b, and c parameters were calculated by fitting the function (Figure 6),

$$R = \left(a^2 + \frac{b^2}{L} + \frac{c^2}{L^2}\right)^{1/2}$$

where *a* refers to light transmission from the scintillator to the photocathode, *b* refers to light production, absorption, photon-electron conversion, and electron amplification, and *c* refers to the contributions of noise originating from the phototube and electronic units. As can be seen, all parameters depend on the characteristics of the detector parts. Therefore, these parameters must be determined for each detector.

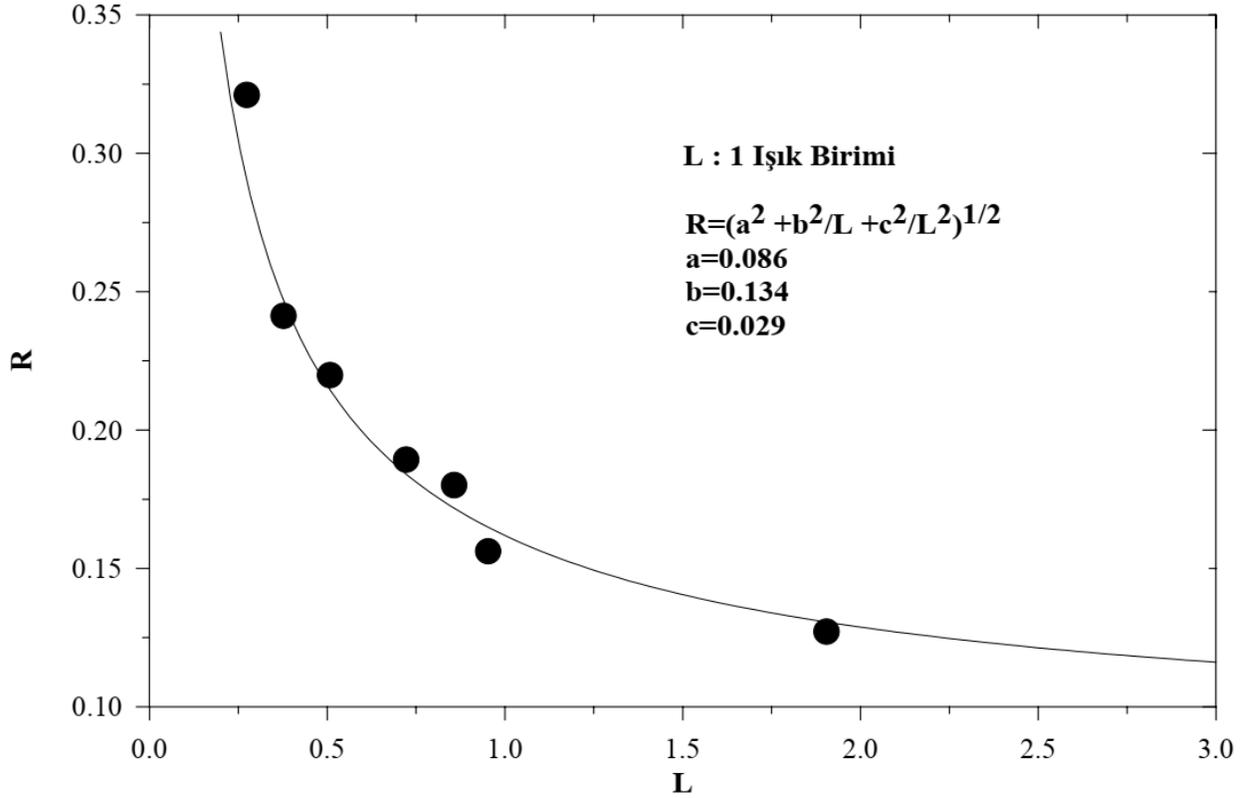

Figure 6. Resolution of the detection system

## 6. Spectrum Unfolding Problem

The pulse height spectrum of the charged protons is in the folded state. This problem of the spectrum unfolding can be expressed in two major ways. First, in the center of mass coordinate system, in the case of isotropic neutron-proton scattering, the density of the recoil protons (proton numbers/MeV.cm³), can be expressed,

$$P(E) = \int_{E'=E}^{\infty} \frac{N(E')\Sigma_n(E')}{E'} dE'$$

where, $N(E')$ represents the fluence of the incident neutron (neutrons/MeV.cm²) and $\Sigma_n(E)$ represents the macroscopic cross-section of Hydrogen. Another way of expressing the problem is

$$P(E) = \int_0^\infty R(E, E')N(E')dE'$$

where, $R(E,E')$ is the response function of the spectroscopy system. The solution of the integral is below. That is

$$N(E) = \frac{-E}{\varepsilon(E)B(E)} \frac{dP(E)}{dE}$$

where $\varepsilon(E)$ represents the energy-dependent efficiency of the detection system, and $B(E)$ is the correction factor caused by multiple scattering and edge effects.

The second solution is to express equation (4.4) in matrix form [13]. That is

$$P_i = \sum_{j=1}^{J} N_j R_{ij} \delta E_j \qquad i = 1, 2, \ldots, I$$

and therefore, the fluence can be expressed as

$$N_j = \sum_{i=1}^{I} R_{ji}^{-1} P_i \qquad j = 1, 2, \ldots, J$$

The solution of this equation can be done by a matrix inversion method. Many computer codes FERD-PC [4], MATUXF [14] have been developed to unfold the recoil proton spectrum. In this study, for the second method, the FERD-PC computer code used the matrix inversion was performed. This code is a kind of version of the FERD code, which is very famous in this field for personal computers. In addition, many features have been added. These are **a)** making the code interactive, **b)** changing the response file easily due to the ease of control of the program, and **c)** calculating the integral neutron flux in the energy range desired by the user. The unfolding process with FERD computer code was first introduced by Burrus and Verbinski [13]. Many codes (FERDOR, SLOP, COOLC) were written later using the method of the FERD code to convert the measured pulse height spectrum to the true neutron spectrum. But the mathematical analysis method is the same as FERD.

To test the unfolding process of the pulsed proton pulse height spectrum, a recoil proton spectrum of the $^{241}$Am-Be source was taken with the PSD of the NE213 detector (Figure 7). As a result of the unfolding of the spectrum with FERD-PC, the neutron spectrum in Figure 8 was obtained. In the obtained spectrum, neutron peaks are seen at 2.2, 3, 5, 7, and 10 MeV. Except for 7 MeV, all measured energy peaks agree with the neutron spectrum obtained due to the work done by Marsh et al. [15]. The peak of 7 MeV appears as a combination of 6.5 and 7.7 MeV peaks. The shape of the energy spectra of $^{241}$Am-Be neutron sources varies according to the source geometry.

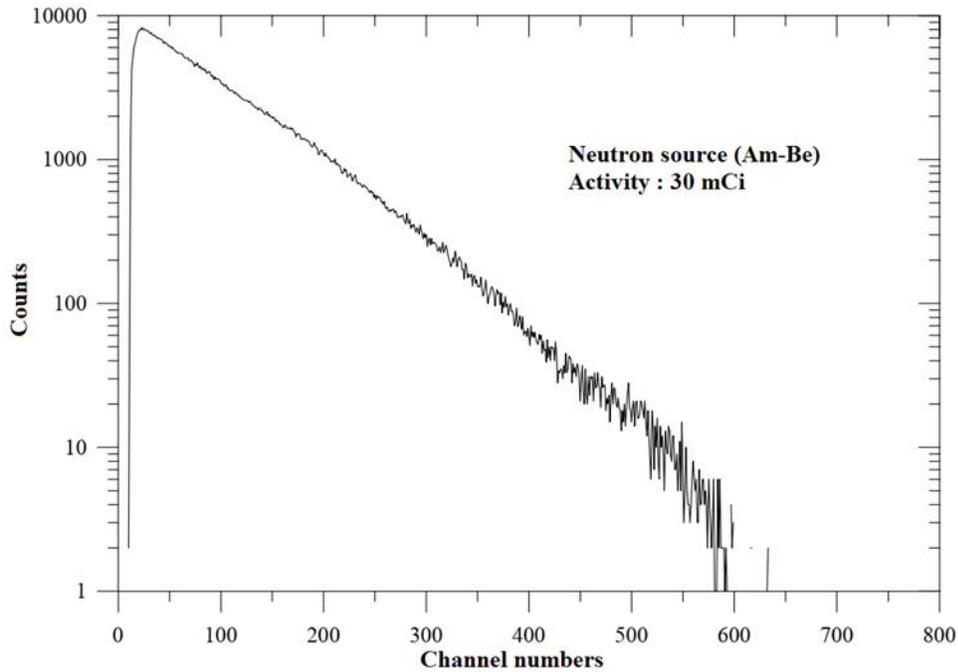

Figure 7. Recoil proton spectrum of $^{241}$Am-Be neutron source

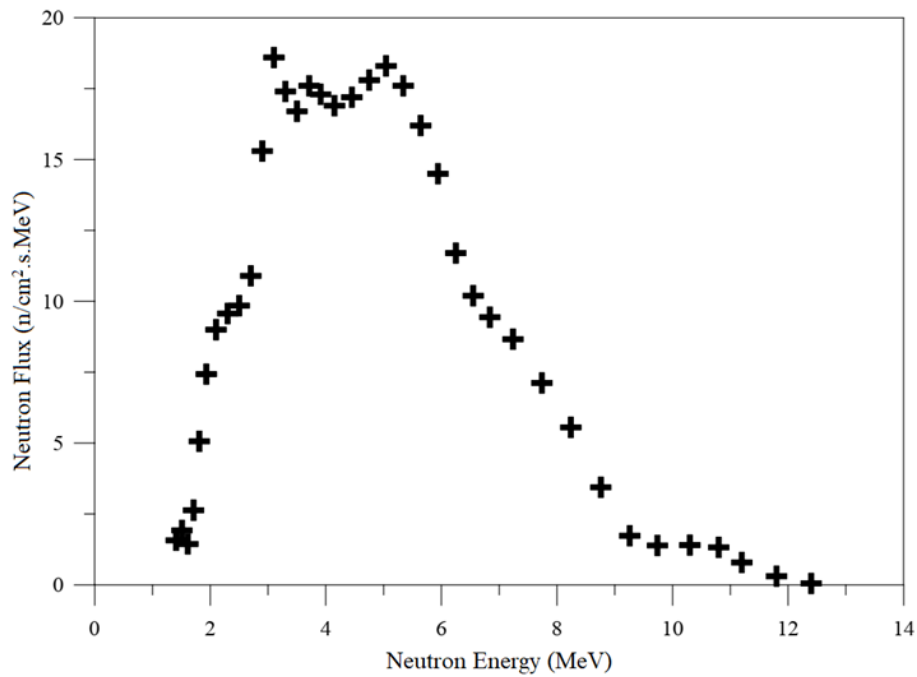

Figure 8. The energy spectrum of the $^{241}$Am-Be neuron source

Thus, it has been seen that the pulse shape discrimination is done correctly in the detection system. The recoil proton spectrum of 14 MeV neutrons (Figure 9) and the neutron spectrum obtained as a result of unfolding this spectrum with FERD-PC are shown in Figure 10. If we look at the spectrum, in addition to the peak seen at 14 MeV, a peak is also seen at 2.5 MeV. This peak belongs to the neutrons resulting from the $d+d\rightarrow{}^3He+n$ reaction that occurs as a result of the build-up of deuterons on the target during irradiation.

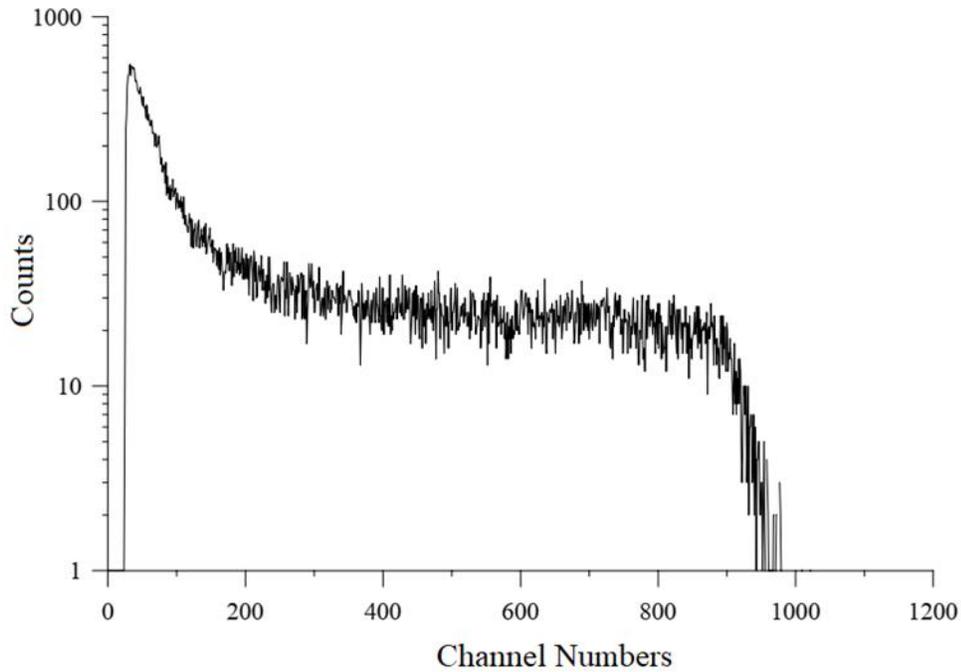

Figure 9. Recoil proton spectrum of 14 MeV neutrons

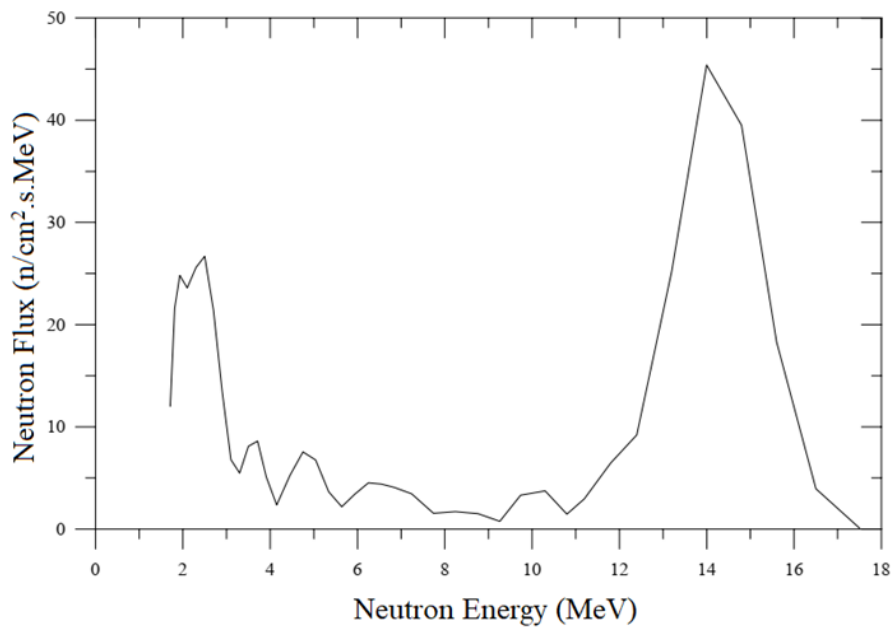

Figure 10. Neutron spectrum of the SAMES J-15 neutron generator

## 7. Conclusion

With the method used in this study, it can be determined the energies and fluxes of neutrons in medium where any neutron/gamma-rays are together (such as reactors, accelerators, and neutron sources). Since the neutron spectrum is very difficult to measure compared to other radiation sources and measurement is made with many different methods, it is necessary to use the fastest method. With this method, In this study, a measurement system composed of NIM modules was used. If a fast digitizer module is used for neutron/gamma-ray separation, the neutron spectrum can be determined online. Thus, energy and flux measurements can be performed in fusion plasma systems quickly.

## Acknowledgements

The author would like to thank Prof. Dr. M. Nizamettin ERDURAN for his support.